\documentstyle[12pt,aaspp4]{article}
\tightenlines
\def\ie{i.e.\ }
\def\eg{e.g.,\ }
\def\etal{et~al.\ }
\def\kms{km s$^{-1}$}
\def\ltsima{$\; \buildrel < \over \sim \;$}
\def\simlt{\lower.5ex\hbox{\ltsima}}
\def\gtsima{$\; \buildrel > \over \sim \;$}
\def\simgt{\lower.5ex\hbox{\gtsima}}
\def\hdb{halo:disk$+$bulge mass ratio\ }
\def\hdbs{halo:disk$+$bulge mass ratios\ }
\def\solar{$_{\sun}$}

\journalid{}{}
\articleid{}{}
\lefthead{Mihos, Dubinski, \& Hernquist}
\righthead{Tidal Tales Two}
\slugcomment{To appear in the {\it Astrophysical Journal}}

\begin{document}

\title{Tidal Tales Two: The Effect of Dark Matter Halos on
Tidal Tail Morphology and Kinematics}

\author{J. Christopher Mihos,\altaffilmark{1,2} John Dubinski,\altaffilmark{3} and Lars Hernquist,\altaffilmark{4,5}}

\altaffiltext{1}{Hubble Fellow}
\altaffiltext{2}{Department of Physics and Astronomy, The Johns Hopkins 
University, hos@pha.jhu.edu}
\altaffiltext{3}{Canadian Institute for Theoretical Astrophysics, dubinski@cita.utoronto.ca}
\altaffiltext{4}{Presidential Faculty Fellow}
\altaffiltext{5}{Board of Studies in Astronomy and Astrophysics, University of California, Santa Cruz, lars@ucolick.org}

\begin{abstract}

We examine the effect of different dark matter halo potentials on the
morphology and kinematics of tidals tails in a merger model of NGC 7252.
We find that models of merging galaxies with low halo masses of
$M_h \sim 4-8 M_{disk+bulge} (M_{db})$ can fit the observed morphology and 
kinematics of the NGC 7252 tails while galaxies with high mass halos 
($M_h \sim 16-32 M_{db}$) fail in this respect.
In high mass models, the deep potential only allows weakly bound disk
material (stars or gas) at $R$ \gtsima 5 disk scale lengths to be ejected in 
tidal tails which tend to fall back onto the parent galaxies before the
final merger. Galaxies with massive, low density halos are somewhat more
successful at ejecting tidal debris during mergers, but still have
difficulties recreating the thin, gas-rich tails observed in NGC 7252.
Our models suggest upper limits for the dark halo masses in the NGC 7252 
progenitor galaxies of roughly $M_h$ \ltsima 10 $M_{db}$. We note, however, 
that our calculations have focused on the rather idealized case of 
the isolated merging of galaxies with distinct dark matter halos; calculations
which employ more realistic (``cosmological'') initial conditions 
are needed to fully explore the use of tidal tails in constraining
dark matter in galaxies.

\end{abstract}


\vfil\eject

\section{Introduction}

While the existence of dark matter halos around galaxies seems well
demonstrated through such diverse kinematic tracers as disk galaxy
rotation curves (\eg Rubin \etal 1982, 1985; Kent 1987), satellite
galaxies and globular clusters (Zaritsky \etal 1989; Zaritsky \& White
1994; Kochanek 1996), and hot gas around ellipticals (\eg Forman,
Jones, \& Tucker 1985), the radial extents and total masses of these
halos remains poorly constrained.  The rotation curves of spiral
galaxies generally probe the mass distribution out to only $\sim$ 10
disk scale lengths, while estimates based on more distant satellites
are statistical in nature, and sensitive to selection effects and
assumptions about orbital kinematics (see, \eg Zaritsky \& White 1994;
Kochanek 1996). Taken together, these lines of argument generally
suggest that galaxies with circular velocities similar to the Milky
Way have halos with masses M$_{\rm halo} \sim 10^{12}$ M$_{\sun}$
and extend beyond $\sim 100$ kpc.

In an attempt to constrain dark matter halos in an independent manner,
Dubinski, Mihos, \& Hernquist (1996; hereafter DMH) showed that the
morphology of tidal tails produced in galaxy collisions depends
sensitively on the potential of the galaxies. The use of tidal debris
to probe dark matter halos was originally proposed by Faber \&
Gallagher (1979), and later emphasized by White (1982) and Negroponte
\& White (1983), who argued that galaxies with massive dark halos
might have difficulty forming long tidals due to their deeper
potential wells.  Barnes (1988) tested these ideas using
self-consistent models and noted a weak anticorrelation between the
masses of the dark halos of the colliding galaxies and the amount of
material ejected in the tidal tails.  However, Barnes used galaxies
with relatively low mass halos (\hdbs of 0, 4, and 8:1) and concluded
that tidal tails are generically easy to produce. Employing halos much
more massive than those used by Barnes, DMH demonstrated that if one
considers halos as massive and as extended as some observations
suggest, the formation of long tidal tails is sharply curtailed.
Given that a number of merging galaxies display long tidal tails (\eg
NGC 4038/39, NGC 7252, the Superantennae), DMH argued that such
galaxies must have \hdbs on the order of 10:1 or less.

DMH's study focussed primarily on the {\it morphology} of tidal tails
produced in various galaxy encounters.  However, the kinematics of
tidal debris may also provide additional constraints which can be
compared directly to observed HI kinematics of merging galaxies (\eg
Hibbard 1994; Hibbard \& Yun 1997). The kinematics of tidal debris
trace the encounter by following trajectories determined in large part
by the orbital energy and interaction geometry.  Hibbard \& Mihos
(1995; hereafter HM) used the morphology and kinematics of the
extended tidal tails around NGC 7252 to reconstruct the dynamical
history of this merger. Their model constrained the orbital geometry
and viewing angle of the encounter as well as the merging timescale,
and predicted future infall rates of material currently populating the
tidal tails. However, HM used a single \hdb of 5.8:1 in their
simulations and did not investigate in any detail the sensitivity of
their results to the internal structure of the merging galaxies.

In what follows, we consider both the morphology and kinematics of
tidal tails formed from collisions of galaxies with various halo
properties, to provide additional constraints on the amount of dark
mass around galaxies as well as to understanding the long term
evolution of the tidal debris.  Our first step is to examine the
kinematics of tidal tails in general, employing models with extended
disks of material to trace the dynamics of the loosely bound material
from which tidal tails are drawn. We then reanalyze NGC 7252,
comparing the morphology and kinematics of the observed tidal tails to
those produced in the models.  Finally, we address the robustness of
the DMH results by considering models with rotating halos and ones
with high mass halos having lower central densities and shallower
potentials.

\section{Models}

The galaxies used in our study are the self-consistent disk/bulge/halo
models developed by Kuijken \& Dubinski (1996). Each galaxy consists
of a disk and bulge with disk to bulge mass ratio 2:1, embedded in a
dark matter halo. In dimensionless units, the disks have a radial
scale length $R_d=1$, circular velocity $v_c(R_d)=1.0$, disk mass
$M_d=0.82$ and bulge mass $M_b=0.42$.  When scaled to the Milky Way,
these values correspond to $R_d=4$ kpc, $v_c=220$ \kms, $M_d=4.4\times
10^{10}$ M$_{\sun}$, and $M_b=2.3\times 10^{10}$ M$_{\sun}$.  Four
dark halo models are used (Models A -- D), varying in their radial
extent and total mass, with \hdbs ranging from 4:1 to 30:1 (see Table
1). The models are chosen to have comparably flat rotation curves
within 5 disk scale lengths, and deviate only at larger radius (see
Figure 1 of DMH).

\begin{table}

\caption{Galaxy Model Properties}

\begin{tabular}{ccccccc}\hline
Model & M$_d$ & M$_b$ & M$_h$ & $R_{\frac{1}{2}}/R_d$ & $R_t/R_d$ & $M_h/M_{db}$\\
 & (1) & (2) & (3) & (4) & (5) & (6)\\ \hline
A & 0.82 & 0.42 & \ 5.2 & \ 3.5 & \ 21.8 & 4 \\
B & 0.82 & 0.42 & \ 9.6 & \ 6.0 & \ 30.1 & 8 \\
C & 0.82 & 0.42 & \ 19.8 & \ 9.1 & \ 44.0 & 16 \\
D & 0.82 & 0.42 & 37.0 & 13.6 & \ 72.8 & 30 \\
E & 1.14 & 0.50 & 32.1 & 30.0 & 115.5 & 20 \\  \hline

\end{tabular}
\tablecomments{(1) disk mass, (2) bulge mass, (3) halo mass, 
(4) half mass radius (5) tidal radius (where density drops to zero)
(6) ratio of halo to disk+bulge mass.}

\end{table}

In addition to a conventional exponential disk, we include a uniform
distribution of test particles at $R=5 - 10 R_d$. These particles do
not contribute to the potential and trace the kinematics of the
loosely bound material at large distances which ends up as tidal
debris. Because the surface density of material at $R>5 R_d$ is small,
self-gravity in the tidal tails is negligible to their overall
kinematic development\footnote{However, self-gravity is important to
the formation of substructure {\it within} the tails (\eg Barnes \&
Hernquist 1992, 1996)}\llap, justifying the use of massless test
particles for this exercise.

Since our goal is to compare the model kinematics to the observed
HI kinematics of NGC 7252, the choice of orbital geometry for the
merging galaxies is largely based on the NGC 7252 model of HM. Disk
geometry is defined by the inclination ($i$) of the disk to the
orbital plane and the argument of periapse ($\omega$) (see Toomre \&
Toomre 1972). In the HM model, the disk which gives rise to the
northwest tail is oriented at ($i,\omega$) = ($-40,0$), while the disk
forming the east tail has an orientation of ($i,\omega$) = ($70,-40$).
HM used a perigalactic separation of $R_p=2.5 R_d$, but noted a
degeneracy between $R_p$ and halo compactness, in that distant mergers
with compact halos merged on similar timescales as close passages with
more diffuse halos (see also Barnes 1992, DMH). Accordingly, we use
two values $R_p = 2$ and $R_p = 4$ to examine this effect. Finally, we
use a zero energy orbit for the encounters, starting each simulation
with the galaxies separated by approximately twice the radius of the halos.

\begin{figure}
\plotone{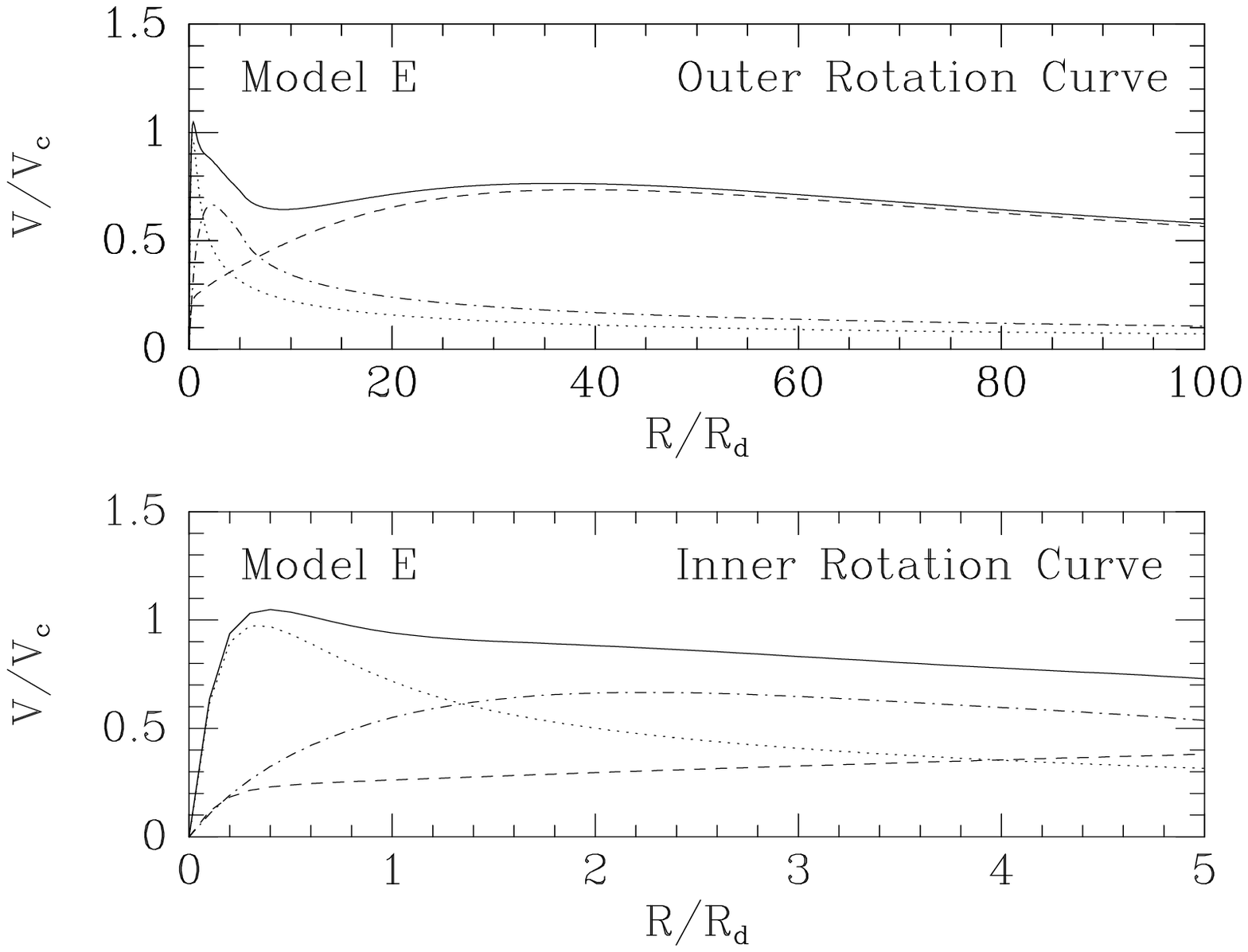}
\vspace{10pt}
\caption{Rotation curves of the inner and outer regions of Model E.
The total mass of this model intermediate to the high mass models C and
D, but the lower central density leads to a more extended, shallower 
potential.  The velocity in the
inner region is $V_c \sim 1.0$ but drops off asymptotically to $V_c \sim
0.7$ at large radii.}
\label{fig-vrot}
\end{figure}

We also follow up attempts by DMH to generate long tidal tails in
galaxies with high mass halos by introducing two new galaxy models.
The first is motivated by recent studies which suggest that the halos
of luminous spirals may have lower circular velocities than that
of the disk (Persic, Salucci \& Stel 1996; Navarro, Frenk \& White 1996). 
The halos in models such as these may still be quite massive, but would 
be much more extended and have a shallower potential. We construct
a new galaxy Model E with mass intermediate to Models C and D, but with a 
shallower potential.  The rotational speed in the disk is $V_c(R_d)=1.0$ 
while at large radii ($R > 30 R_d$), $V_c \approx 0.7$ and is flat out to 
$100 R_d$ (Figure \ref{fig-vrot}).  The mass of the disk and bulge are $M_d=1.13$, $M_b = 0.5$,
slightly larger than in Models A-D to compensate for the lower halo mass
within the disk, ensuring the inner rotation curves (at $R<5R_d$) are
similar to models A -- D. The total halo mass in Model E is $M_h=32$,
intermediate to Models C and D, and the halo extends to R=115, or
nearly 0.5 Mpc (see Table 1).

The second new model incorporates halo rotation into Model D,
accomplished by giving all halo particles the same sign of $z$ angular
momentum.  The resulting dimensionless spin parameter is
$\lambda=0.20$.  Halo rotation has been shown to increase the strength
of dynamical friction between a halo and a precessing disk (Nelson \&
Tremaine 1995).  Halo rotation might, therefore, lead to more
resonances between the halo and passing companion (much like the
resonances in the disk which give rise to tidal tails) which could
hasten merging and lead to the development of longer tidal tails
than in the Model D mergers with non-rotating halos.  The spin of
these halos is significantly larger than expected from cosmological
arguments, which give $\lambda=0.05$ (\eg Warren \etal 1992), so they 
represent an extreme of this effect if it is present.

Aside from Model E, a total of 160,000 particles were used to
represent each galaxy (320,000 particles per merger simulation):
40,000 in the exponential disks, 20,000 in the bulges, 60,000 in the
halos, and 40,000 in the extended test particle disks.  In the Model E
galaxies, 80,000 particles were used in the disk and 100,000 in the
halos.  All models were run using a parallel treecode (Dubinski 1996)
on the T3D at the Pittsburgh Supercomputing Center.  A leapfrog
timestep of $\Delta t=0.1$ was used, resulting in energy conservation
to better than 2\%.

We note that our calculations focus on the merging process in
isolated environments -- there are no neighboring companions,
nor is there any ambient potential well in which the galaxies merge
(as would be found in cluster or group environments). As such, these 
models still represent an idealized version of merging and tidal tail 
formation, and calculations with more realistic (and complex) merger 
dynamics will be necessary to fully explore the use of tidal tails 
as a means to constrain dark matter distributions in galaxies.

\section{Results}

\subsection{Kinematics of Tidal Debris}

We first examine the global kinematics of the tidal tails produced in
the encounters. Unfortunately, choosing the most appropriate time to
compare the different models is not straightforward. The
overall dynamics of both the encounter and the tails depend on the
galaxy halos and so the merging times and the times at which the tails
achieve their maximum lengths are quite different in the various runs.
Consequently, if the models are compared at the same time following
the beginning of each simulation, the systems would be in different
dynamical states.  For simplicity in making the comparison, we choose
to ``observe'' the models one half-mass rotation period after the
galaxies have merged\footnote{We define ``merged" here to mean the
point at which the center of mass kinetic energy of the quarter-most
bound particles in the central bulges is zero.} in each calculation, 
and focus on
collisions between galaxy Models A--D with $R_p=4$. The morphological
and kinematic trends observed in the closer $R_p=2$ mergers are qualitatively 
similar to those in the $R_p=4$ mergers described below. 

\begin{figure}
\plotone{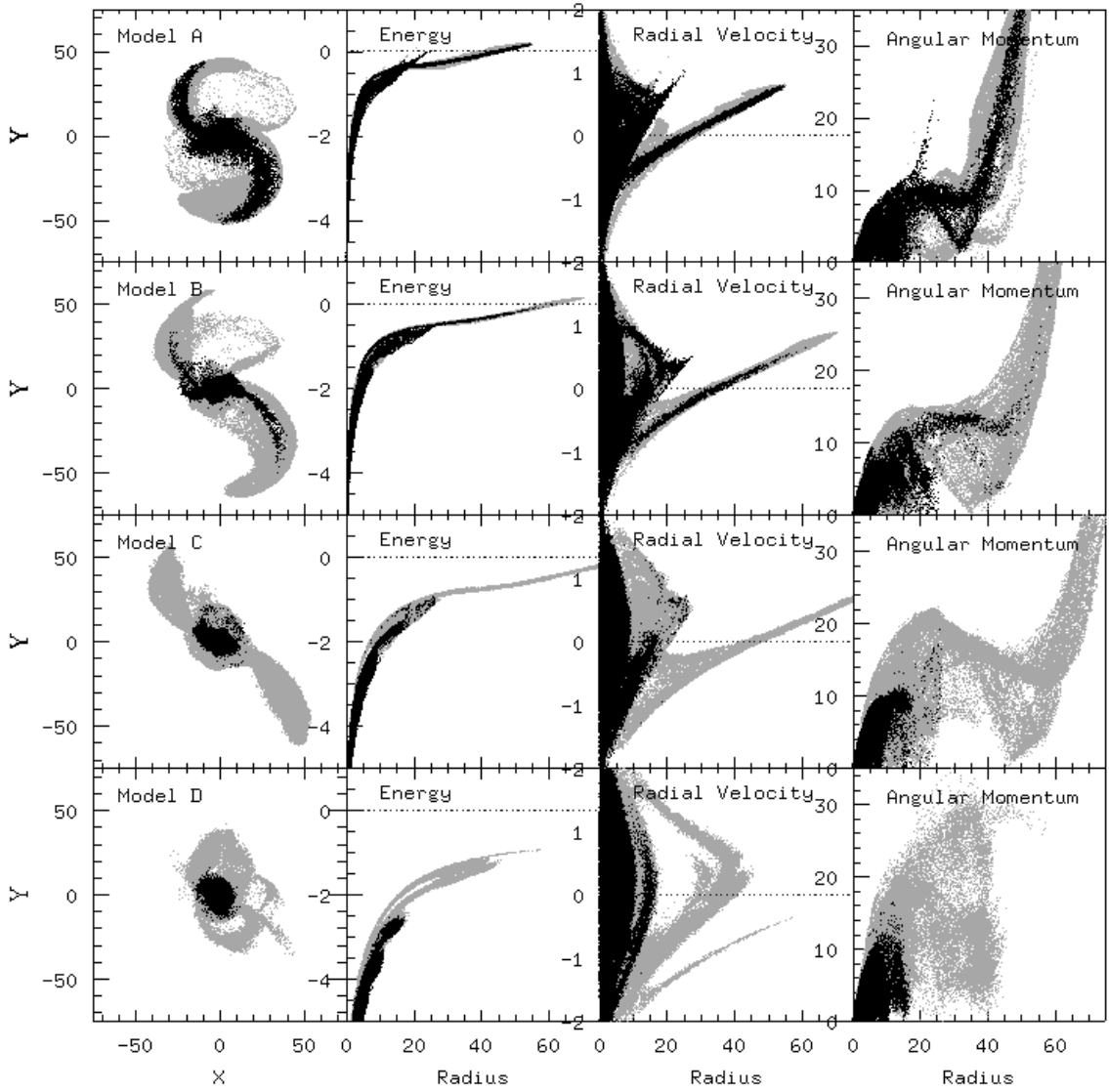}
\vspace{10pt}
\caption{Morphology and kinematics of the $R_p=4$ merger models, viewed
one half-mass rotation period after the galaxies have merged. From
left, the panels show the morphology of the tails (projected onto the
orbital plane), and the energy, radial velocity (with respect to the
central merger remnant), and angular momentum of the  tidal material
as a function of radius.}
\label{fig-tailkin}
\end{figure}

Figure \ref{fig-tailkin} shows the morphology of the tidal tails formed in each
encounter (projected onto the orbital plane), along with the energy, 
radial velocity, and angular momentum as a function of radius along
the tidal tails.  The self-gravitating particles which form the inner
exponential disk (at $R_{init}<5R_d$) are shown in black, while the
outer, flat distribution of test particles is shown in grey.

Two cautionary notes are in order when interpreting the outer test
particles. First, since they are initially distributed with constant
surface density, the number of particles increases as $r^2$.  As a
result, the morphology of the outer tails in Figure \ref{fig-tailkin} is dominated by
particles at very large radius. In real galaxies the mass distribution
is typically dropping with distance, so that tidal debris may be
significantly more limited in extent than shown in Figure \ref{fig-tailkin}.  Second,
while the outer parts of galaxies are usually HI dominated, the test
particles in our simulations are collisionless and can pass through
the galaxies and/or merger remnant without experiencing shocks and
dissipation. For example, material seen leading the tidal tails in
Model A, or the ``third tail'' in Model B, comes from particles which
passed through the galaxies shortly after the initial collision; gas
would likely not survive on such trajectories (\eg Hernquist \& Barnes
1991; Hernquist \& Weil 1992). However, most of the material in the
extended tails does not suffer such orbit crossing, indicating that
the test particles should do a good job of tracing the overall
kinematics of extended tidal debris.

The influence of the dark matter halos on the morphology of the tidal
tails formed from the inner material is very similar to that described
by DMH.  Galaxies with low mass halos produce massive, curving tails.
As we consider encounters involving galaxies with increasing halo
mass, the tails become straighter and more anemic, until for the
highest halo mass explored, the tails have nearly disappeared.
However, the outer test particles trace the tidal material to larger
distances than do the particles in the exponential disk, and show that
the tidal debris can be more complex than suggested by DMH.

In Model A, the tails are quite long, and are comprised of particles
from both the inner and outer disk. Indeed, it is perhaps surprising
that material from the inner disk extends as far out in the tails as
does the outer disk material. The latter broadens the tails, and
traces the curvature of the tails to larger distances, but is not more
extended. The tails are still mostly expanding, with only debris near
the base of the tails falling back inwards.  The distribution of
energy and angular momentum also shows the inner and outer material
are well mixed in radius, and the fact that the binding energy along
the tail runs smoothly through zero indicates that the outermost
material in the tails will continue to expand as the remnant evolves,
even as the inner, bound material falls back.

Examining mergers of galaxies with increasing halo mass, the amount of
inner disk material in the tails decreases, and is found mostly at
smaller distances with lower binding energy. In Model B, the inner
disk material still traces the tails, but unlike Model A, none of
these particles are unbound.  The situation is even more extreme in
Model C, where the tails consist entirely of outer test particles,
while the inner disk material has fallen back into the remnant to form
shells (cf. Hernquist \& Spergel 1992; HM).  The outer disk material
is still expanding in Model C, but the turnaround radius has slowly
marched outwards so that particles within 40$R_d$ are already falling
back towards the remnant. None of the inner or outer disk material is
unbound in Model C, although the particles with binding energies close
to zero will remain at large distances for many Gyrs.

Finally, in the merger of two Model D galaxies, remnant tidal tails are
not found in {\it either} the inner or outer disk material. Tails
are launched shortly after the galaxies first collide, but the
particles comprising these features are tightly bound to the galaxies
and fall back onto the galaxies before they actually merge, as noted
by DMH.  Consequently, once the merger is complete, the tidal debris
has already been accreted by the remnant, surrounding it in the form
of loops and shells.

As indicated by Figure \ref{fig-tailkin}, the radial extent of the disks can affect
the lengths of the tidal tails, an effect not considered by DMH.  The
more loosely bound material in the outer disks is readily expelled
into the tails.  Figure \ref{fig-rinitfin} shows the radii in the original disks from
which material in the tails was drawn. Low mass mergers extract
material from deep within the colliding galaxies, sending it to large
distances in the tails.  For larger halo masses, the inner disk
material is more tightly bound to the host galaxy, and the tails are
formed from material initially further out.  For Model C, only
particles with initial radii larger than $5-6 R_d$ contribute to the
tails, while in Model D, the extended loops are comprised only of
material from radii greater than $7 R_d$.

The fact that tidal tails are formed from material initially at
different locations within the progenitors indicates that the amount
of mass (or, alternatively, stellar luminosity) comprising the tails
may depend sensitively on the asymptotic structure of the colliding
disks.  To quantify this finding, we can assign masses to the outer
test particles {\it ex post facto} for various adopted initial mass
distributions, and derive the total mass of the ensuing tidal tails.
The choice for the initial mass distribution depends on the component
of interest: stellar disks follow the exponential density profile
continued from the inner disk, while HI disks in galaxies generally
follow a flatter profile, with more mass at large radii. To span a
range of plausible outcomes, we choose four surface density profiles
for setting the masses of the test particles: $\Sigma=$constant,
$\Sigma \sim r^{-1}$, $\Sigma \sim \exp(-r/2R_d)$, and $\Sigma \sim
\exp(-r/R_d)$.  The cumulative mass profiles of the tidal debris
derived by this procedure are shown in Figure \ref{fig-massdist}.

For an outer disk with constant surface density, the tidal debris is
quite massive, even for the Model D mergers.  However, this is an
extreme limiting case, and probably does not reflect the actual HI
profiles of disk galaxies. For mass distributions more typical of
extended HI disks, 10--20\% of the material ends up in the tidal tails, 
and, for lower mass halos, much of this material is expelled to great distances 
from the remnant (and is still expanding outwards). In contrast, for 
similar mass distributions, the Model D merger has only a few percent of 
this material in the tidal debris; most of the particles remain
at small distances, tightly bound to the remnant. For a pure
exponential disk (comparable to the stellar mass distribution in
galaxies), the mass in the tails is $\sim$ 15--20\% of the {\it total}
disk mass for Model A mergers, $\sim$ 10\% for Model B mergers, and 
$\sim$ 5\% for Model C mergers.  The Model D mergers, with no true tails, 
contain only a few percent of any exponentially distributed material in 
their tidal debris.

We can compare these values to the HI and stellar luminosity observed
in the tidal debris of NGC 7252.  Hibbard \etal (1994) find $\sim
2\times 10^9$ M\solar\ of HI in the tidal tails, and a blue luminosity
for the tails of $\sim 3\times 10^9$ L\solar, or $\sim$ 7\% of the
total blue luminosity of the system. These values suggest that NGC
7252 is best described by our Model B encounters -- the amount of
``starlight'' in the Model A mergers is too large for the observed
blue luminosity of the tails in NGC 7252, while Models C and D have
tails that are too anemic, by comparison.

\subsection{NGC 7252 Comparison}

We now compare the simulations directly to the morphological and
kinematic properties of NGC 7252 from Hibbard \etal (1994). For each
model, we attempt to find the observing geometry and time which best
fit the HI data, although in some cases that ``best fit'' may not be
ideal. Our goals are to estimate how unique a given solution is,
once variations in halo properties are taken into account, and to see
if additional constraints can be placed on the halo properties of the
galaxies which collided to form the NGC 7252 system.

We begin by eliminating models which are obviously discrepant. As
noted earlier, the tidal debris in Model D collisions has already
fallen back into the remnant by the time galaxies merge; extended
tidal tails do not persist in this case. Mergers of Model C galaxies
are somewhat more difficult to dismiss.  Several arguments, however,
make these simulations a poor fit to NGC 7252.  The tidal tails in the
calculation are comprised of material located initially only outside
6$R_d$; with an exponential distribution of starlight, this would put
$\sim$ 0.5\% of the stellar luminosity in the tails, an order of
magnitude less than the 7\% of NGC 7252 starlight (L$_B$) in NGC 7252
actually detected in the tails (Hibbard \etal 1994; see also \S 4 below). 
Furthermore, because the tails drawn from the Model C galaxies come entirely 
from loosely bound outer disk material, they are highly warped, making it
difficult to associate them with the relatively thin tidal features
observed in NGC 7252.  Finally, the Model C tails possess very little
of the large-scale curvature needed to reproduce the structure of the
northwestern HI tail in NGC 7252. For these reasons, we also reject
the Model C mergers as being good matches to NGC 7252.

We focus now on mergers of Model A and B galaxies, which have \hdbs of
4 and 8, respectively.\footnote{Note that the HM model of NGC 7252
employed galaxies with a \hdb of 5.8, intermediate to our Models A and
B.} For each simulation, we examined the remnant at two times: one
immediately following the final coalescence of the progenitors and
another after the remnant has evolved for a few dynamical times. In
comparing the model remnant to the observations, three things, in
particular, determined the subjective quality of the fit: the
curvature of the northwest tail, the kinematic gradients along the
tails, and the straightness of the eastern tail. The latter constraint
was the most difficult to match, because of the strong warping of the
outer disks. If much of the material in the eastern tail of NGC 7252
came from loosely bound gas, the disk giving rise to the eastern tail
must have been more closely aligned to the orbital plane than the
$i=40$\arcdeg\ value of the HM model.

For brevity, we show each model only at the time when it best matches
the HI observations of NGC 7252. The HI data from Hibbard \etal (1994) 
is shown in Figure \ref{fig-7252data}, in the form of the ``clean
components'' of the VLA data cube (see HM for details). 
Figure \ref{fig-7252fit} shows the morphology
and projected kinematics of the four models, and can be
directly compared to Figure \ref{fig-7252data}.
The best fit time in each case proved to be near $t=120$, or 70 time
units ($\sim 1$ Gyr) after the galaxies first collided.
For the Model A and the $R_p=2$ Model B simulations, all of which
resulted in mergers soon after first passage, the remnant is more
evolved than that for the $R_p=4$ Model B calculation, where the
remnant is somewhat younger.

Figure \ref{fig-7252fit} clearly shows both the difficulty in obtaining an ideal
fit, and the degeneracies which complicate matching the simulations to
observations.  Nonetheless, some trends are apparent with both $R_p$
and halo mass which help to constrain the parameters. One important
diagnostic is the curvature of the NW tail, and it appears that the
simulations here bracket a best fit -- the Model A mergers have NW
tails which are too curved, while the corresponding tails in the Model
B mergers are too straight.  The curvature of the tails is also
manifested by a hook-shaped feature in the kinematic plots (most
noticeable in the $Y-V_r$ projection). These features were {\it not}
reproduced in the dynamical model of HM, because that model did not
include any of the extended outer disk material which makes up the
hook.  The hook is most noticeable in the Model A mergers, and is less
apparent in the Model B mergers, again suggesting that the two models
bracket a best fit.

As noted above, it proved very difficult to reproduce the linearity of
the eastern tail. This problem was hinted at in the HM model; their
tail had a slight southern curvature where it joined to the merger
remnant. The present models, which include material at much larger
initial distances than the HM model, emphasize this problem -- to
varying degrees, all the models have eastern tails which do not extend
radially from the remnant. However, the problem is less severe for the
Model A mergers when only the portions of the tails which arise from
material inside $R=5$ (shown in black in Figure \ref{fig-7252fit}) are considered.
This material is not so strongly warped out of the disk plane,
resulting in a more linear eastern tail. This solution is not
applicable to the Model B mergers, where the tidal tails arise almost
exclusively from the loosely bound outer disk material.

Taken together, the various models indicate that NGC 7252 is best fit
by mergers of progenitors with halo masses in the range of $M_h
\sim 4-8\times M_{db}$. Unfortunately, this estimate is not tightly
constrained. Because the curvature of the tidal tails is determined in
large part by the orbit of the merging galaxies, there is a tradeoff
between halo mass and perigalacticon making similar solutions possible
for different choices of the orbital geometry and structure of the
galaxies.  For example, Models B2 and B4 show that for the same halo,
wider encounters produce tails that are more curved. As a result, it
may be difficult to distinguish between close collisions of low mass
models and wider collisions involving more massive galaxies. This
argument cannot be taken to extremes, however, as very distant
encounters would not have had sufficient time to merge before the time
set by the dynamical state of the tidal tails.  For example, distant
Model C mergers would have neither the amount of stellar mass in the
tails nor the dynamical age necessary for a satisfactory fit to NGC
7252.  But within the stated mass estimate given here, many
satisfactory solutions will exist -- a single, unique solution is
probably unattainable.

\subsection{Variant Halo Models}

The Model C and D collisions presented here and in DMH consistently demonstrate
the difficultly in producing long stellar tidal tails from merging galaxies
with massive halos.  However, it is slightly misleading to refer to the halo 
mass as the only parameter which controls these differences, since it is
really the shape and gradient of the galactic potential which
determines the evolution.  One could add mass to a halo with a shallow
potential simply by extending the halo to a greater distance and
reducing the central density. 

To examine the evolution of tidal tails in mergers of galaxies with 
massive, low density halos, we set up a collision between two
Model E galaxies with zero energy orbits and pericentric distances of
$R_p=2.0$ and $4.0 R_d$.  Figure \ref{fig-modelE.starsgas} shows the 
evolution of the Model E $R_p=4.0$ collision.  At first encounter, material 
is ejected into tidal features, but like the Model C and D galaxies the debris 
is limited in extent, and largely formed from material beyond 
$\sim 5 R_d$.  Because of the lowered halo density, dynamical friction is weaker
than in the fiducial mergers, and the merging timescale is very long:
260 time units, or $\sim$ 4.5 Gyr. As a result, the initially-ejected
material has ample time to fall back into the galaxies well before 
they merge. Upon the second passage preceding the merger, this material is 
re-ejected as the diffuse tidal tails visible in the final remnant.

At first glance, the final stage of the Model E merger looks similar to
the low mass mergers (Models A and B), in the sense that it does display 
extended tidal tails. In detail, however, several problems remain.
At intermediate stages, the tidal debris remains wrapped around the
galaxies, unlike the long tidal features shown by galaxy pairs such
as NGC 4676 (The Mice) or Arp 295 (see, \eg Hibbard 1995). Once the
galaxies have merged, the extended debris is very diffuse, unlike
the thin stellar tails of NGC 7252.
Furthermore, the tidal tail in the model E merger is made exclusively
from material in the outer disk of (collisionless) test particles, which
are initially ejected during the first passage, fall back into the galaxy,
and are re-ejected during the final merging. Were this material gas-rich,
it would likely not follow this collisionless evolution, but instead
suffer significant orbit crossing and strong dissipation, making
the formation of long tidal tails during the final merging very difficult.
Because the extended tidal tails observed in NGC 7252 are very gas-rich,
it seems difficult to describe NGC 7252 by a merger of Model E progenitors.

Another alternative model for the halo is one which includes rotation.
Rotating dark halos can potentially increase the dynamical braking
during a galaxy collision through a stronger resonant coupling between
the orbits of the galaxies and the particles making up the halos.
This effect is seen in simulations of satellite accretion where
satellites quickly sink to the center of a galaxy once they settle
into the equatorial plane of the disk, e.g. Quinn, Hernquist \&
Fullagar (1993), Walker, Mihos \& Hernquist (1996).  Nelson \&
Tremaine (1995) have also shown that rotating halos can change the
strength and sign of dynamical friction in the context of tilted disks
precessing in flattened halos, although their analysis applies equally
to any external perturbations.  With this motivation, we examined a
collision between two Model C galaxies with rotating halos
and compared it to the nonrotating halo cases above.  
 
A comparison of the trajectories of the colliding galaxies with and
without halo rotation exhibit few significant differences.  After
their encounter, the galaxies in the two simulations were separated by
nearly the same distances and merged at virtually the same time,
suggesting that halo rotation in this case has little effect on
merging.  Not surprisingly, the resulting tidal debris is essentially
unchanged from that in the non-rotating model C mergers.  Our study is, 
however, not exhaustive, so it is still possible that halo rotation could 
have an effect for different galaxy orientations and orbital geometries 
(perhaps in nearly coplanar, direct encounters), but it had little effect 
on the evolution of the system thought to have produced NGC 7252.

Finally, we note that several recent studies suggest that dark matter
halos may be significantly non-spherical. Dark matter halos which form in 
cosmological N-body simulations are strongly triaxial, with minor-to-major 
axis ratios of $\sim$ 1:2 (\eg Dubinski \& Carlberg 1991; Warren \etal 
1992; Steinmetz \& Muller 1995). Observations of polar ring galaxies (Sackett
\etal 1994) and warped gas disks (Olling 1996) hint at even flatter shapes
for dark matter halos. Although still subject to great uncertainties,
these results raise the question of how dependant our results are on
the assumption of spherical halos. To answer this, we emphasize that the 
structure of the tidal tails at the time of merging is governed
by two factors: (1) the gradient in the potential well, and (2) the
merging timescale. The timescale for merging is set largely by the
total mass, which determines the encounter velocity irrespective
of the halo shape. While the potential gradient is more sensitive to
the halo shape, we point out that isopotential contours are significantly
rounder than isodensity contours; to make any significant impact on our 
results, halos must be {\it extremely} flattened (\ie disk-like). 
The usual kinematic disk instabilities (\eg Ostriker \& Peebles 1973)
make such ``disky'' dark matter models highly untenable.

\section{Summary and Discussion}

The models presented here expand on the work of DMH and HM in two
respects.  First, we have followed the evolution of material initially
located at very large distances in the progenitor galaxies, allowing
us to examine the detailed kinematics and morphology of this loosely
bound material.  In simulations involving mergers of galaxies with
increasing halo mass, the tidal debris is drawn primarily from
particles located at increasingly large radii within the progenitors,
and more of this material remains bound to the merger remnant.  The
tidal tails which form immediately in mergers involving very massive
halos quickly fall back into the galaxies, so that they are no longer
visible by the time the galaxies merge.  These models reinforce the
claim of DMH that observed merger remnants with long tidal tails must
have formed from progenitors with relatively small \hdbs\llap.
Second, we have explored a variety of halo models in an attempt to
reproduce the observed tidal tail morphology and kinematics of NGC
7252, placing some constraints on the dark matter distribution around
merging galaxies.  Again, the observations are best fit using mergers
of galaxies with \hdbs in the range of 4--8. However, we find that
precise estimates are difficult because of degeneracies in the
solution, as originally suggested by HM.

\begin{figure}
\plotone{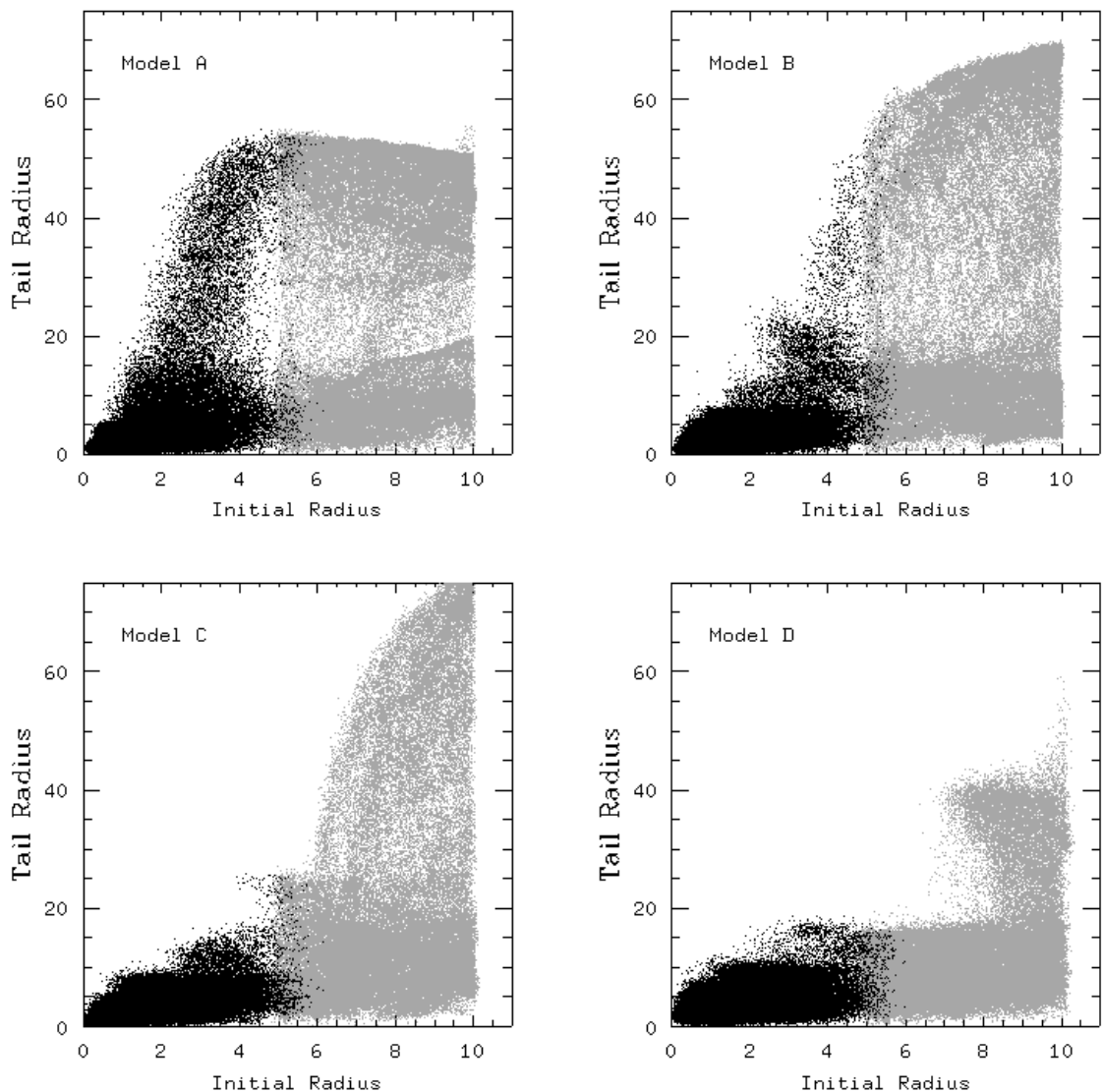}
\vspace{10pt}
\caption{Tail radius versus initial radius for the models shown
in Figure \ref{fig-tailkin}. ``Tail radius'' is defined as the radial distance of
material in the tidal tails one half-mass rotation period after the 
galaxies have merged. Mergers involving galaxies with low mass halos 
draw material from deep within the progenitor disks, while the debris 
formed in mergers of galaxies with massive halos is comprised only of
loosely bound material from the extreme outer portions of the disks.}
\label{fig-rinitfin}
\end{figure}

\begin{figure}
\plotone{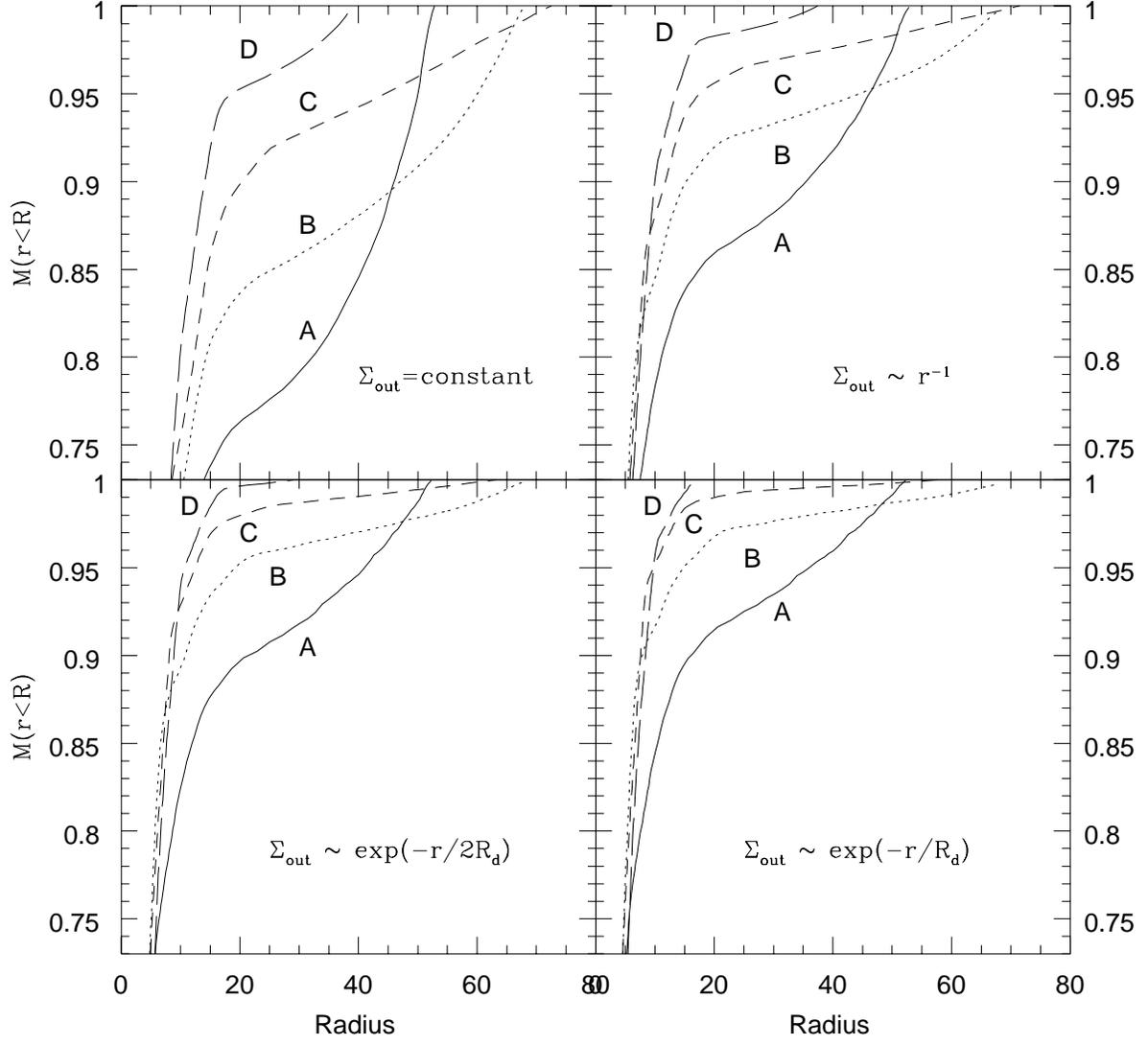}
\vspace{10pt}
\caption{Cumulative mass distribution for the mergers shown in Figure
\ref{fig-tailkin}, under the assumption of varying initial mass distributions for the
test particle material at $R_{init} > 5 R_D$.}
\label{fig-massdist}
\end{figure}

\begin{figure}
\plotone{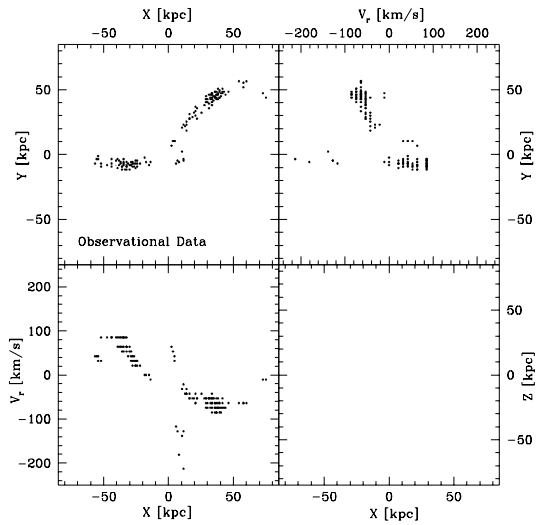}
\vspace{10pt}
\caption{HI morphology and kinematics of NGC 7252 (from Hibbard \etal
1994; Hibbard \& Mihos 1995). The ``clean components'' of the HI data
cube are shown (see Hibbard \& Mihos 1995 for details). The panels
show the HI morphology (upper left), $Y-V_r$ position-velocity diagram
(upper right), and $V_Z-X$ position-velocity diagram (lower left).}
\label{fig-7252data}
\end{figure}

\begin{figure}
\plotone{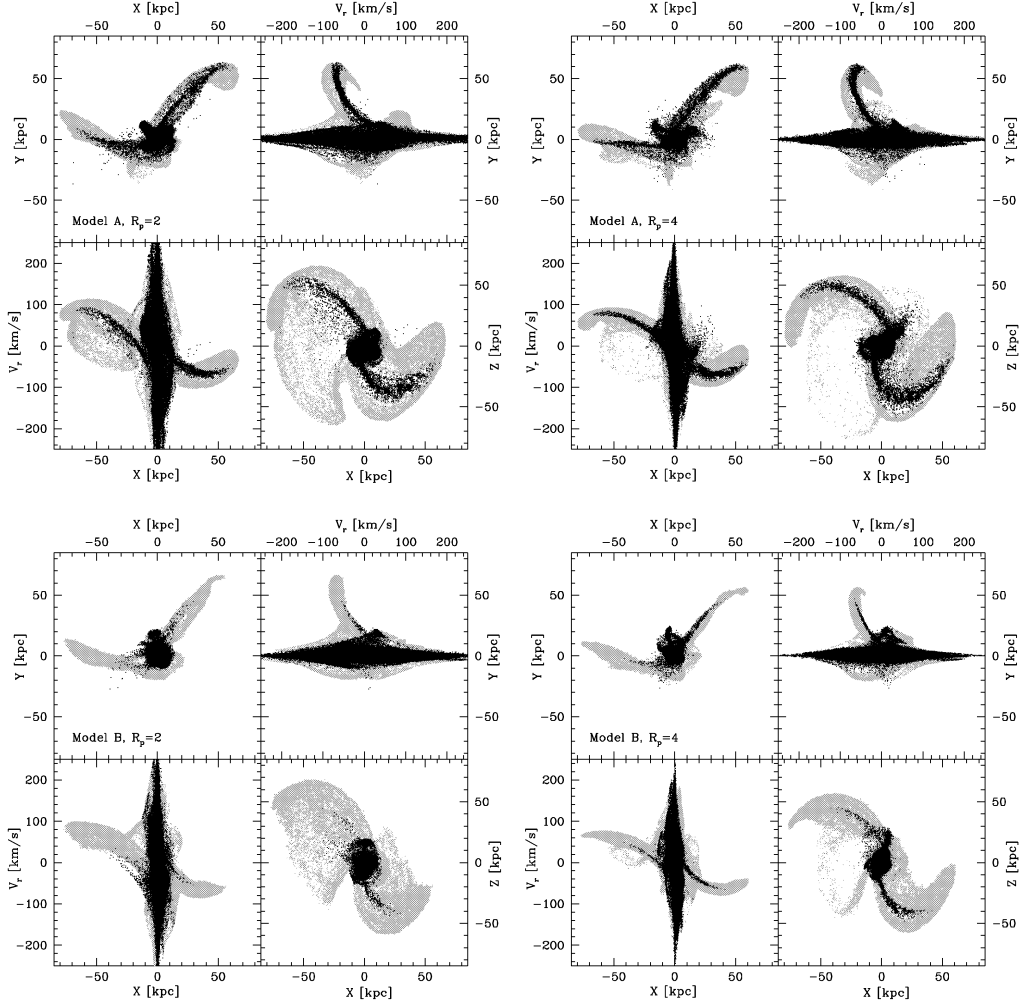}
\vspace{10pt}
\caption{Projection of 4 $N$-body models for comparison with the 
observed HI morphology and kinematics of NGC 7252 (\ref{fig-7252data}). Each 
subframe compares the morphology (upper left), $Y-V_r$ position-velocity diagram
(upper right), and $V_Z-X$ position-velocity diagram (lower left), and
also shows a ``top view'' of model remnant (lower right).}
\label{fig-7252fit}
\end{figure}

\begin{figure}
\plotone{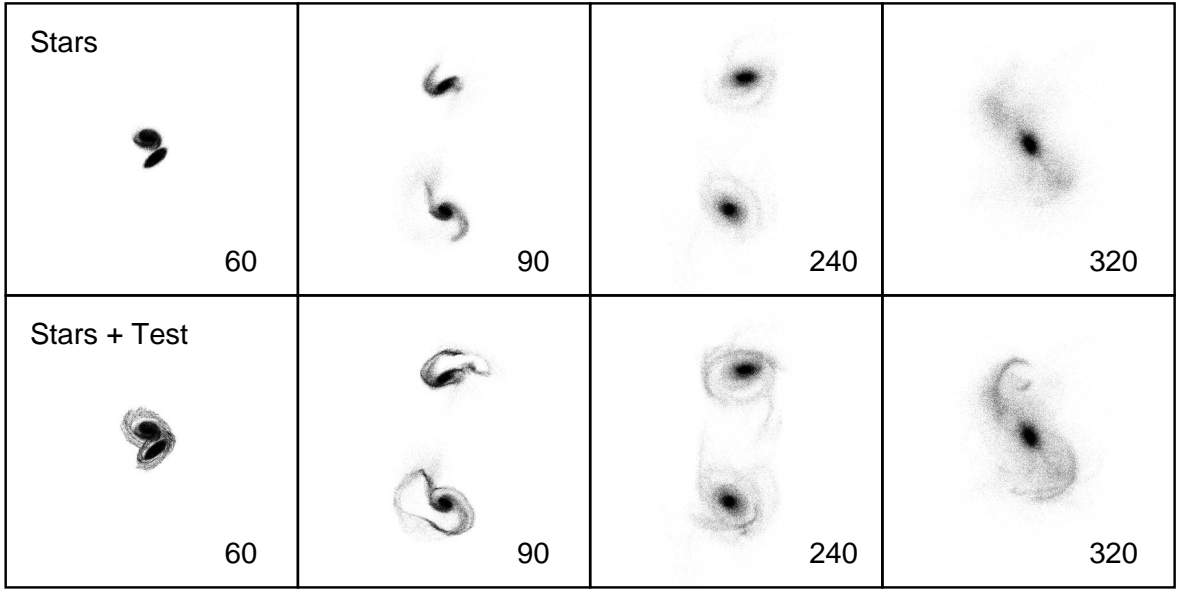}
\vspace{10pt}
\caption{Evolution of the Model E interaction with $R_p=4.0$
in the orbital plane. Top: Exponential disk particles ($R<5R_d$)
only.  Bottom: Exponential disk plus outer test particles. Each box is 100 
units wide (400 kpc) and time is shown in each frame (unit time equals 
17 Myr).
The dynamical breaking is weak in these models because of the smaller
central halo density and so the galaxies pass by each other quickly.
The short duration of the encounter leads to the modest excitation of some
tidal arms which fall back onto the galaxy before the second encounter.
Fairly long tidal tails are ejected during the final merger although they
are composed of material beyond $R \simgt 5 R_d$.}
\label{fig-modelE.starsgas}
\end{figure}

The tendency of tail material to be drawn from larger initial radii with 
increasing halo mass (Figure \ref{fig-rinitfin}) has implications for recent 
QSO absorption line studies. Because of abundance gradients in galactic disks, 
the ability of galaxy interactions to expel metal-rich material to large 
distances will be sensitive to the mass distributions of dark matter halos.  
Accordingly, the metallicity of tidal tails may be used as another 
constraint on the masses of galaxy halos. Absorption lines produced by
tidal debris from intervening galaxies have been identified in several
QSO spectra (\eg Sargent \& Steidel 1990; Norman \etal 1996); while
metallicity estimates are uncertain, if such systems prove to be
reasonably metal-rich, it would support the idea that galaxy halos may be less
massive than other observational estimates. This argument is similar
to the suggestion that low-mass galaxies are more able to eject metal-rich
material into the IGM through starburst-driven superwinds (\eg Heckman, Armus, 
\& Miley 1990); in this case, however, the energy involved in expelling 
metal-rich material comes from tidal encounters rather than starburst winds.

With the extended disk models showing that galaxies with more massive halos
may eject significant amounts of extended HI into tidal tails, the possibility 
arises that subsequent star formation could convert this gas into stars and 
produce the long {\it optical} tidal tails observed in some merger remnants.
However, to match observed tidal tails, which contain as much as 10--20\% of 
the blue luminosity of merging galaxies, this star formation must be
prodigious. For example, if the optical light in the
tidal tails of NGC 7252 were to come from stars formed {\it in situ},
then $\sim$ 80\% of the gas in the tails must have been converted into
young stars at several M$_{\sun}$ yr$^{-1}$ to reproduce the total
blue luminosity and observed (remaining) gas content of the tails.
While some star formation is observed in tidal tails, it typically
occurs in a few star forming clumps rather than being smoothly
distributed, and at much lower rates. Furthermore, the observed colors
of tidal tails are more representative of material stripped from the
inner disks of galaxies, rather than young stellar populations
(Schombert \etal 1990).

We have also investigated interactions using galaxies containing halo
models with different internal kinematics and mass distributions.
Maximally rotating halos ($\lambda = 0.20$) have no discernible effect
on the evolution of a Model C merger and so the amounts of rotation
inferred in halos from cosmological arguments ($\lambda=0.05$) are
unlikely to be important for determining the evolution of merging
galaxies.  Mergers of galaxies with high mass, extended halos (Model E) 
are able to eject more material into tidal debris, due to their
shallower potential wells. However, the tidal debris is very
diffuse and suffers significant orbit crossing, making it difficult to 
identify with the gas-rich tidal features in objects like NGC 7252. However, 
our models have only examined one representation
of a low density halo model and a more systematic study of the
effects of low density halos is warranted, especially in light of
observational (Casertano and van Gorkom 1991; Persic \etal 1996) and
theoretical (Navarro \etal 1996) results which suggest that the most
luminous spiral galaxies may have declining rotation curves.

The models described here address many of the loopholes left open by
DMH, supporting the conclusion that long tidal tails are a signature 
of compact, low-mass halos in the progenitors to a merger.  Nonetheless, 
one limitation of the models still remains: their rather idealistic initial 
conditions -- galaxies with individual, distinct dark matter halos merging 
in the absence of any background potential.  Given recent cosmological 
simulations which show that galaxy halos often merge before their luminous 
galaxies do (\eg Katz, Hernquist, \& Weinberg 1992), our simulations may 
be an oversimplified version of galaxy mergers. In fact, dynamical
friction against a background dark matter distribution may hasten
merging, allowing massive galaxies to merge before their tidal tails
have fallen back into the galaxies. The development of tidal tails
will also be affected by the interaction between {\it three} potential
wells (two galactic and one background); the structure and kinematics of 
the resultant tidal features is difficult to assess without detailed 
modeling.  The next consistency check on our results would therefore be 
to examine mergers in a more ``cosmological'' setting, in which galaxies
merge in a more diffuse ``sea'' of dark matter. 

In principle, the statistics of tidal tails could be used to infer the
properties of dark matter halos. In practice, however, this may be
difficult to achieve.  While long tidal tails suggest low mass halos,
the converse may not necessarily be true -- the lack of observed tidal
tails may have been the result of an unequal mass merger, an
unfavorable orbital geometry (\ie a retrograde merger), unsuitable
progenitors (ellipticals or S0's), or rapid fading in surface
brightness due to kinematic evolution of the tails (\eg Mihos 1995).
Furthermore, sample selection would be fraught with bias -- as mergers
are generally identified through the presence of tidal debris, care
would need to be taken to ensure the sample would not be skewed
towards low mass systems with obvious tidal tails.

While a statistical constraint on the dark matter content of galaxies
using tidal tails may be problematic, the implications for individual
systems seem more clear. For merging galaxies such as NGC 7252, the
Antennae, and the Superantennae, the presence of long tidal tails is
difficult to reconcile with massive dark matter halos, unless perhaps 
the halos are very extended and diffuse. While most
kinematic probes of the mass distribution in galaxies (\ie rotation
curves, satellite kinematics) yield lower limits on halo masses, the
results described here suggest some of the first {\it upper limits} on
the dark matter content of galaxies.  As such, it is of immediate
interest to test these concepts using both numerical simulation and
detailed observational studies of the morphology, kinematics, and
metallicity of tidal tails.  As coherent kinematic tracers at the
largest radius, tidal tails may yet unveil the dark matter halos in
which galaxies live.

\acknowledgments

We thank John Hibbard for many lively discussions and for providing
the HI data for comparison with the simulations. We also thank the
Aspen Center for Physics, where the first version of this paper was
drafted. This work was supported in part by the Pittsburgh
Supercomputing Center and the NSF under Grant ASC 93--18185 and the
Presidential Faculty Fellows Program. JCM is supported by NASA through
a Hubble Fellowship grant \#~HF-01074.01-94A awarded by the Space
Telescope Science Institute, which is operated by the Association of
University for Research in Astronomy, Inc., for NASA under contract
NAS 5-26555.


\clearpage

\end{document}